\documentstyle[aps,graphicx,tighten,preprint]{revtex}
\begin{document}
\title{Contradiction of Quantum Mechanics with Local Hidden Variables for 
Continuous Variable Quadrature Phase Amplitude Measurements \\ }
 \vskip 1 truecm
\author{A. Gilchrist\\ }

\address{Physics Department, University of Waikato, Hamilton, New Zealand\\ }  	  
\author{P. Deuar and M. D. Reid\\ }

\address{Physics Department, University of Queensland, Brisbane, Australia\\ }  	  
\date{\today}
\maketitle
\vskip 1 truecm
\begin{abstract}
We demonstrate a contradiction of quantum mechanics with local hidden variable 
theories for continuous variable quadrature phase amplitude 
(``position'' and ``momentum'') measurements, by way of a violation 
of a Bell inequality. For any 
quantum state, this contradiction is lost for situations   
where the quadrature phase amplitude results are always macroscopically 
distinct. 
We show that for optical realisations of this experiment,  
where one uses 
homodyne detection techniques to perform the quadrature phase amplitude 
measurement, one has an amplification prior to detection, so that 
macroscopic fields are incident on photodiode detectors. The 
high efficiencies of such detectors may open a way for a loophole-free 
test of local hidden variable theories.
\end{abstract}
\narrowtext
\vskip 0.5 truecm
In 1935 Einstein, Podolsky and Rosen $^{\cite{1}}$ presented an argument 
for the incompleteness of quantum mechanics. The argument was based on 
the validity of two premises: no action-at-a-distance (locality) 
and realism. Bell $^{\cite{2}}$ later showed that the predictions of  
quantum mechanics are incompatible with the premises of local 
realism (or local hidden variable theories). 
Experiments $^{\cite{3}}$ based on 
Bell's result support quantum mechanics, indicating the failure of local hidden 
variable theories.

One feature appears 
characteristic of all the contradictions of quantum mechanics with local 
hidden variables studied to date. The measurements considered have  
 discrete outcomes, for example being measurements of spin or photon number. 
By this we mean specifically that the eigenvalues of the appropriate
 system hermitian operator, which represents the measurement in quantum 
 mechanics, are discrete.

In this paper we show how the 
predictions of quantum mechanics are in disagreement with those of local hidden variable theories 
for a situation 
involving continuous quadrature phase amplitude (``position'' 
and ``momentum'') measurements. By this we mean
 that the quantum predictions for the probability of obtaining results 
 $x$ and $p$ for position and momentum (and various linear combinations of these 
 coordinates) cannot be predicted by any local 
 hidden variable theory. This is of fundamental interest since the 
 original argument $^{\cite{1}}$ of Einstein, Podolsky and Rosen 
 was given in terms of position and momentum 
 measurements. The original state considered by Einstein, Podolsky and 
 Rosen, and that produced 
 experimentally in the realisation by Ou et al $^{\cite{4}}$ of this 
 argument, 
 gives probability 
 distributions for $x$ and $p$ completely compatible with a local hidden 
 variable theory.

 Second we suggest a new macroscopic aspect 
 to the proposed failure of local hidden variable theories for the 
 case where one uses optical homodyne detection to 
realise the quadrature phase amplitude measurement $^{\cite{4,5}}$. 
The homodyne detection method employs a second ``local-oscillator'' field which 
combines with 
the original field to provide an amplification prior to photodetection. 
In these experiments then 
large field fluxes fall incident on highly efficient photodiode detectors, in 
dramatic contrast to the former photon-counting experiments. A microscopic 
resolution (in absolute terms) of this incident photon number 
is not necessary to obtain the violations with local hidden variables. 
This is in contrast to many previously cited macroscopic proposals 
$^{\cite{6}}$ for which it appears necessary 
to resolve the incident photon number 
to absolute precision in order to show a contradiction with local hidden 
variable theories. 

The high efficiency of detectors available in this more macroscopic detection 
regime may provide a way to test local hidden 
variables without the use of auxiliary assumptions $^{\cite{2,7}}$ 
which have weakened the 
conclusions of the former photon counting measurements.  
This high intensity limit has not 
been indicated by previous works $^{\cite{8}}$ which showed 
contradiction of quantum 
mechanics with local hidden variables using homodyne detection, 
since these analyses were restricted to a very low intensity of ``local 
oscillator'' field.

We consider the following two-mode entangled quantum 
superposition state $^{[9, 10]}$: 
\begin{equation}
	|\Psi> = N \int_{0}^{2\pi} 
	|r_{0}e^{i\varsigma}>_{A}
	|r_{0}e^{-i\varsigma}>_{B} d\varsigma 
	\label{eqn:circle state}
\end{equation}
Here $N$ is a normalisation coefficient. The 
$|\alpha>_A$, where $\alpha = r_{0}e^{i\varsigma}$, 
is a coherent state of fixed amplitude $r_0=|\alpha|$ but varying phase 
$\varsigma$, for a system
at a location $A$. Similarly $|\beta>_B$, where $\beta = r_{0}e^{-i\varsigma}$ 
and $r_0=|\beta|$, is a coherent state of fixed amplitude but varying phase 
for a second system at a location  
$B$, spatially separated from $A$. 
The quantum state (1) is potentially generated,  
from vacuum fields, in the steady state by 
nondegenerate parametric oscillation $^{\cite{10}}$ 
as modelled by the following 
Hamiltonian, in which coupled signal-idler loss 
dominates over linear single-photon loss. 
\begin{equation}
   H = i\hbar E ( \hat a_1^\dagger \hat b_1^\dagger - \hat a_1 \hat b_1) + 
   \hat a_1 \hat b_1 \hat \Gamma ^\dagger + 
   \hat a_1^\dagger \hat b_1^\dagger \hat \Gamma
	\label{eqn:hamiltonian}
\end{equation}
The 
$\hat a_1^\dagger$ and $\hat a_1$, and $\hat b_1^\dagger$ and $\hat b_1$, are 
the usual boson creation and destruction operators for  
the two spatially separated systems (for example field modes) at 
locations $A$, and $B$, respectively. 
In many optical systems the $\hat a_1$ and $\hat b_1$ are 
referred to as the signal and idler 
fields respectively.
Here E represents a coherent driving parametric term which generates signal-idler pairs, while 
$\hat \Gamma$ represents reservoir systems which give rise to the coupled 
signal-idler loss. The Hamiltonian preserves the 
signal-idler photon number difference operator 
$\hat a_1^\dagger \hat a_1$-$\hat b_1^\dagger \hat b_1$, 
of which the quantum state (1)  is an eigenstate, with eigenvalue zero.
We note the analogy here to the single mode ``even'' and ``odd'' coherent 
superposition states $N_{\pm}^{1/2}(|\alpha> \pm |-\alpha>)$ (where 
$\alpha$ is real and $N_{\pm}^{-1}=2(1 \pm exp(-2|\alpha|^2)$) which are 
generated by the degenerate form (put $\hat a_1=\hat b_1$) 
of the Hamiltonian (2). These states for large $\alpha$ are analogous to the famous 
``Schrodinger-cat'' states $^{\cite{11}}$ and have been recently experimentally generated 
$^{\cite{12}}$. 

Consider the experimental situation depicted in Figure 1. 
Measurements are made of the field quadrature phase 
amplitudes $X_{\theta}^A$ at 
location $A$, and $X_{\phi}^B$ at location $B$. Here we define  
$X_{\theta}^A=\hat a_1exp(-i\theta)+\hat a_1^\dagger exp(i\theta)$; 
and $X_{\phi}^B=\hat b_1exp(-i\phi)+\hat b_1^\dagger exp(i\phi)$. 
Where our system is a harmonic oscillator, 
we note that the angle choices $\theta$ (or $\phi$) equal to zero and $\pi /2$ will 
correspond to position and momentum measurements respectively. 
The result for the amplitude measurement $X_{\theta}^A$ is a 
continuous variable which we denote by $x$. Similarly 
the result of the measurement $X_{\phi}^B$ is a continuous variable 
denoted by $y$. 

We formulate a Bell inequality test for the experiment 
depicted by making the simplest possible binary classification of the 
continuous results $x$ and $y$ of the measurements. We classify the result of the 
measurement to be $+1$ if the quadrature phase result $x$ (or $y$) is greater or 
equal to zero, and $-1$ otherwise.  With many measurements we build up the 
following probability distributions: $P_{+}^{A}(\theta)$ 
for obtaining a positive value of 
$x$; $P_{+}^{B}(\phi)$ for obtaining a positive $y$; and 
$P_{++}^{AB}(\theta,\phi)$ the joint probability of obtaining a positive 
result in both $x$ and $y$. 
  
If we now postulate the existence of a local hidden variable theory, we can 
write the probabilities $P_{\theta,\phi}(x,y)$ 
for getting a result $x$ and $y$ respectively upon the 
simultaneous measurements $X_{\theta}^A$ 
and $X_{\phi}^B$ in terms of the hidden variables $\lambda$ as follows.
\begin{eqnarray}
P_{\theta,\phi}(x,y)= \int \rho(\lambda) \quad p_{x}^A(\theta, \lambda ) 
p_{y}^B(\phi, \lambda )\quad d\lambda  
\label{eqn hidden variables}
\end{eqnarray}
The $\rho(\lambda)$ is the probability distribution for the hidden 
variable state denoted by {$\lambda$}, while $p_{x}^A(\theta, \lambda )$ 
is the probability of obtaining a result $x$ upon measurement at $A$ of 
$X_{\theta}^A$, given the hidden variable state {$\lambda$}. The  
 $p_{y}^B(\phi, \lambda )$ is defined similarly for the results and 
 measurement at $B$. The independence of $p_{x}^B(\theta, \lambda )$ on 
 $\phi$, and $p_{y}^B(\phi, \lambda )$ on $\theta$ is a consequence of the 
 locality assumption, that the measurement at $A$ cannot be influenced by 
 the experimenter's choice of parameter $\phi$ at the location $B$ (and 
 vice versa)$^{\cite{13}}$ . It follows that the final 
 measured probabilities $P_{++}^{AB}(\theta,\phi)$ can be 
 written in a similar form 
 \begin{eqnarray}
P_{++}^{AB}(\theta ,\phi )= \int \rho(\lambda) \quad p_{+}^A(\theta, \lambda ) 
p_{+}^B(\phi, \lambda )\quad d\lambda  
\label{eqn hidden variable2}
\end{eqnarray}
where we have simply  
$ p_{+}^A(\theta, \lambda ) = \int_{x\geq 0} p_{x}^A(\theta, \lambda ) dx$, and 
similarly for $p_{+}^B(\phi, \lambda )$. It is well known that 
one can now deduce $^{\cite{2}}$ the following
 ``strong'' Bell-Clauser-Horne inequality.
  \begin{eqnarray}
 S={{P_{++}^{AB}(\theta,\phi)-P_{++}^{AB}(\theta,\phi')+P_{++}^{AB}(\theta',\phi)
 +P_{++}^{AB}(\theta',\phi')}\over{P_{+}^{A}(\theta')+P_{+}^{B}(\phi)}} \leq 1
	\label{eqnbell}
\end{eqnarray}

The calculation of the quantum prediction for $S$ for the quantum 
state (1) is straightforward.  
We note certain properties of 
the distribution $P_{++}^{AB}(\theta,\phi)$: 
it is a function only of the angle sum $\chi=\theta + 
\phi$ so we can abbreviate $P_{++}^{AB}(\theta,\phi)=P_{++}^{AB}(\chi)$;  
$P_{++}^{AB}(\chi)=P_{++}^{AB}(-\chi)$; and the marginals satisfy  
$P_{+}^{A}(\theta)=P_{+}^{B}(\phi)=0.5$. Results for $S$ are shown in 
Figure 2, for the choice of measurement angles 
$ \theta +\phi= \theta'+\phi'= -(\theta'+\phi)=\pi/4,\theta 
+\phi'=3\pi/4$ (for example put $\theta=0, \phi=-\pi/4, \theta'=\pi/2$ and 
$\phi'=-3\pi/4$). This choice allows the simplification 
$S=3P_{++}^{AB}(\pi/4)-P_{++}^{AB}(3 \pi/4)$. It can 
be shown that for small $r_0$ (less than about $1.5$) this angle choice 
maximises $S$.

Violations of the Bell inequality, and hence contradiction with the 
predictions of local hidden variables, are indicated for 
    $0.96 \lesssim r_0 \lesssim 1.41$,
the maximum violation of $ S \approx 1.0157\pm 0.001$ being 
around $r_0 \approx 1.1$. This is a substantially smaller violation than 
obtained in the discrete case (where $ S \approx 1.2$) of spin 
measurements, considered originally by Bell. The choice of Bell 
inequality and quantum state to give a violation may not be optimal, but 
nevertheless the possibility of a contradiction of quantum mechanics with 
local hidden variables is established.

We note that the violations are lost at large coherent 
amplitudes $r_0$. In this limit the quantum probability 
distributions for $x$ and $y$ 
show two widely separated 
peaks (as indicated by Figure 3), the $+1$ and $-1$ results of the  
measurement then corresponding to macroscopically distinct outcomes, 
resembling the ``alive'' and ``dead'' states of the 
``Schrodinger cat'' $^{\cite{11}}$. 
We obtain asymptotic (large $r_0$) analytical 
forms for the probability distributions which allow a complete search for 
all angles. Results indicate no violations of the Bell inequality (5) 
possible.

In fact it can be demonstrated 
that, for any quantum state, there is no 
incompatibility with local hidden variables 
 for the case where the quadrature phase amplitude results  
 $x$ and $y$ only take on values which 
 are macroscopically distinct. In this case, the addition of a 
noise term of order 
the standard quantum limit (this corresponds to a variance 
$\Delta ^2 x=1$) 
to the result of quadrature phase amplitude measurement will 
not alter the $+1$ or $-1$ classification of the  
result. Yet it can be shown that the quantum predictions for the 
results of such a noisy experiment 
are given by the quantum Wigner function 
$W(x_{0}^A, x_{\pi/2}^A, x_{0}^B, x_{\pi/2}^B)$ for the state (1), convoluted 
by the gaussian noise term  
$(1/4{\pi}^2)\exp \left(-\left[{x_{0}^A}^2+ {x_{\pi/2}^A}^2+{x_{0}^B}^2 
+{x_{\pi/2}^B}^2\right]/2 \right)$. 
This new Wigner function is always positive  $^{\cite{14}}$ and 
can then act as a local hidden variable 
theory which gives all the predictions in the truly macroscopic ``dead'' 
or ``alive'' classification limit.

 An examination however of the 
 homodyne method of 
 measurement of the quadrature phase amplitudes reveals a 
 macroscopic aspect to the experiment proposed here for optical fields.
The optical realisation $^{\cite{4,5}}$ of the quadrature phase amplitude 
measurement (see Figure 1) involves local oscillator fields at $A$ and $B$, which we 
designate by 
the boson operators $\hat a_2$ and $\hat b_2$ respectively. 
The measurement of 
$X_{\theta}^A$  
proceeds when the local oscillator field at $A$ is combined with the 
field $\hat a_1$ 
using a beam splitter to form two combined fields  $\hat{c}_\pm=\left(\hat 
a_2\pm \hat a_1 \exp (-i\theta) \right)/\sqrt{2}$.  
A variable phase shift $\theta$ allows choice of the particular 
observable to be measured. Direct detection, using two 
photodetectors, of the intensities  
$\hat c_\pm^\dagger \hat c_\pm$ of the combined fields and 
subtraction of the two resulting photocurrents results in measurement of 
$ I_D=\hat c_+^\dagger \hat c_+-\hat c_-^\dagger \hat c_-=s_{\theta}^A$ 
where $s_{\theta}^A =
\hat a_2^\dagger  \hat a_1 \exp (-i\theta)+\hat a_2  \hat a_1^\dagger \exp 
(i\theta)$. In the limit where the 
local oscillator fields are very intense one may replace the boson 
operators $\hat a_2$ and $\hat b_2$ by classical amplitudes $E_A$ and $E_B$ 
respectively. Assuming $E_A=E_B=E$ where $E$ is real, we see that 
$s_{\theta}^A=EX_{\theta}^A$. The $X_{\phi}^B$ are 
measured similarly to $X_{\theta}^A$ using a second beam splitter (to give 
fields $\hat{d}_\pm=\left(\hat 
b_2\pm \hat b_1 \exp (-i\phi ) \right)/\sqrt{2}$) and pair of 
photodetectors, at location $B$.

  The important point is that the local oscillator acts as an amplifier 
  prior to detection, 
  the operators $s_{\theta}^A, \hat c_\pm^\dagger \hat c_\pm$ and 
  $\hat d_\pm^\dagger \hat d_\pm$ 
 being photon number operators which have a macroscopic  
 scaling 
 in the very intense local oscillator limit $^{\cite{15}}$. 
 Thus in these experiments 
 large intensities fall incident on the photodetectors, and it is 
 not necessary to determine these photon numbers with a microscopic 
 uncertainty in 
 order to arrive at the conclusion that local hidden variable theories are
  invalid $^{\cite{16}}$. This is 
 in contrast with the previous photon counting experiments, and also many 
 previous macroscopic proposals for which it appears that an absolute 
 resolution of the incident photon number is necessary in order to show 
 failure of local hidden variables.    
 Our result then opens possibilities for testing quantum mechanics against local hidden 
 variable theories in a loophole-free way using very efficient 
 photodiode detectors.

\begin{figure}%
\includegraphics[scale=0.6]{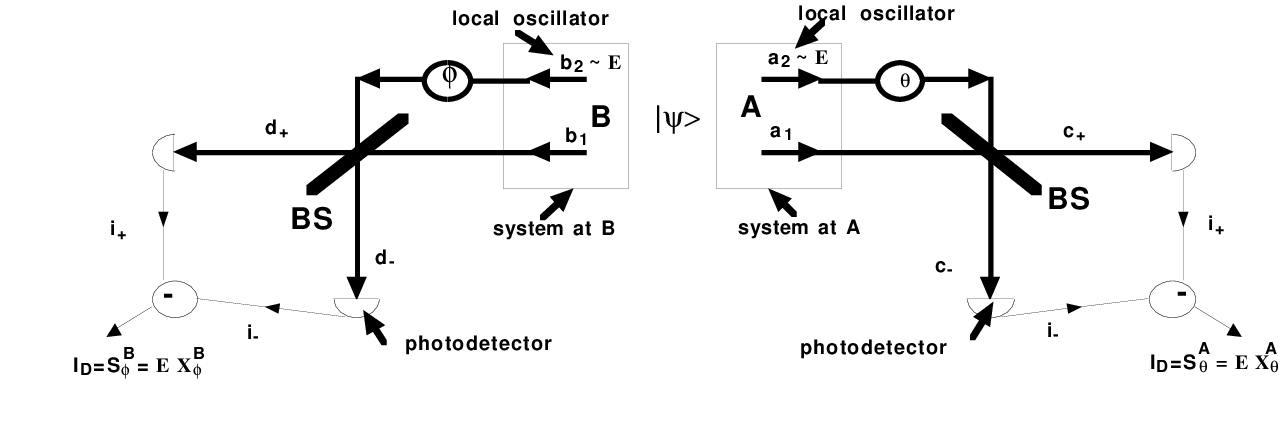}
 \caption
{Schematic representation of a test of the Bell's inequality. 
 Balanced homodyne detection allows 
measurement of the quadrature phase amplitudes $X_{\theta}^A$ 
and $X_{\phi}^B$. }%
\label{fig1}
\end{figure}

\begin{figure}
\includegraphics[scale=0.9]{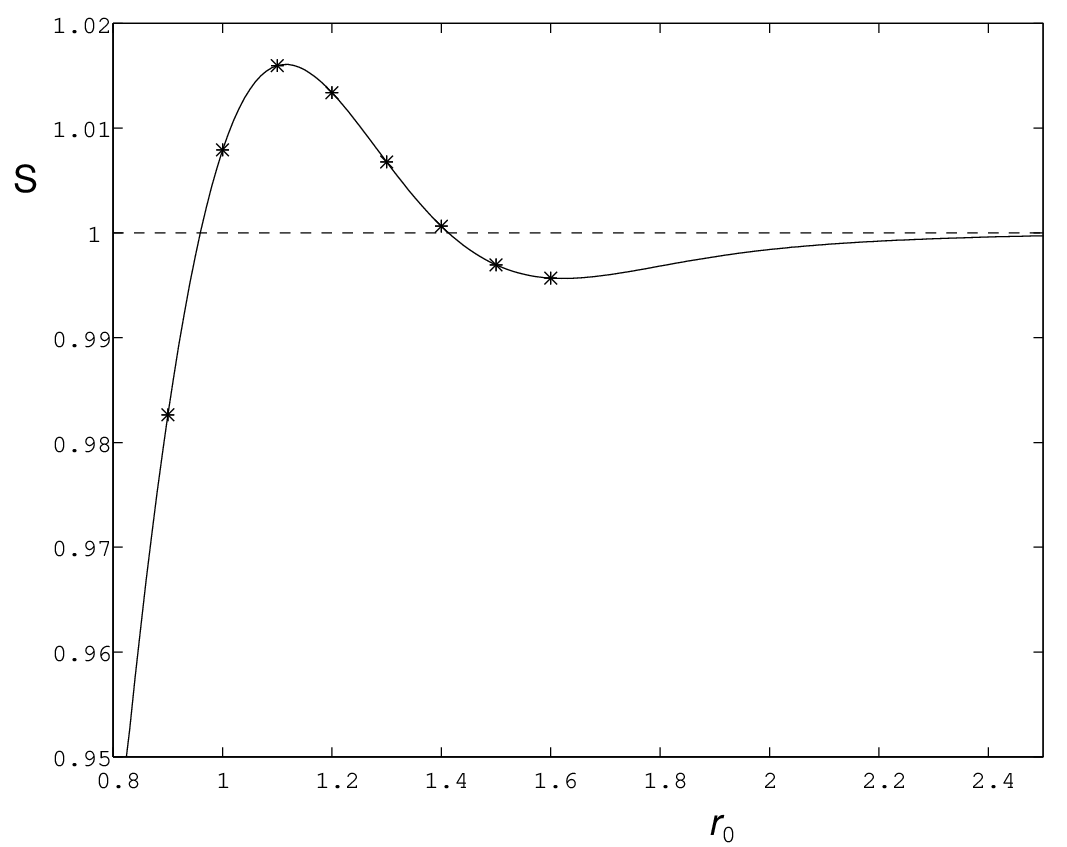}
\caption
{%
Plot of $S$ versus $r_{0}$, for the angle values indicated in the text.}% 
\label{fig2}
\end{figure}

\begin{figure}
\includegraphics[scale=0.7]{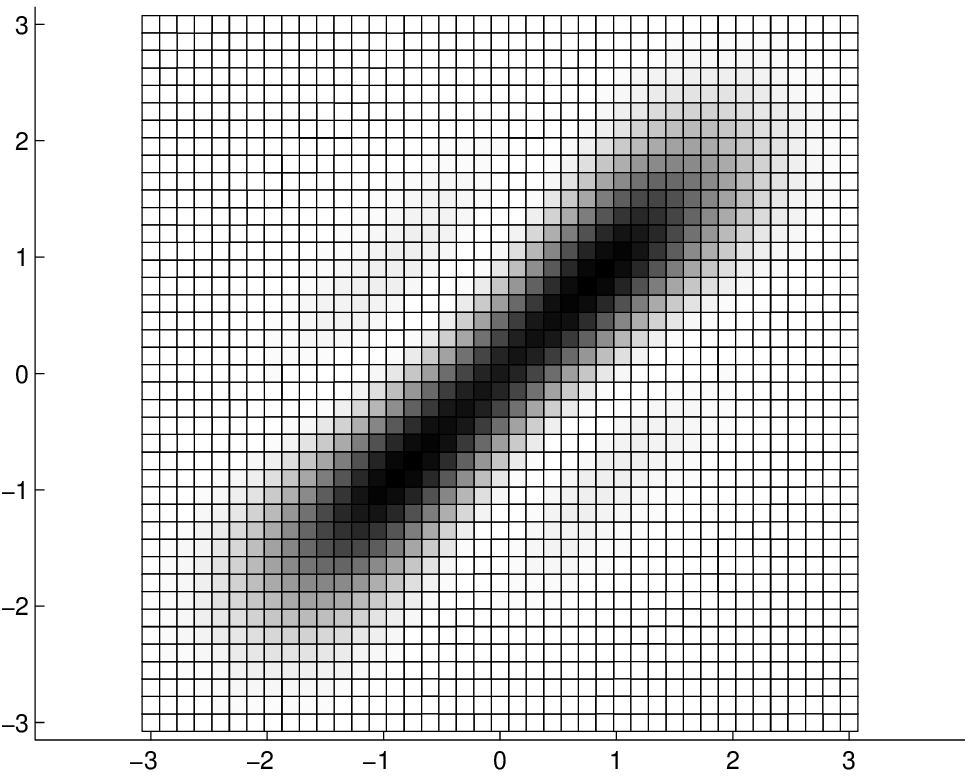}
\includegraphics[scale=0.7]{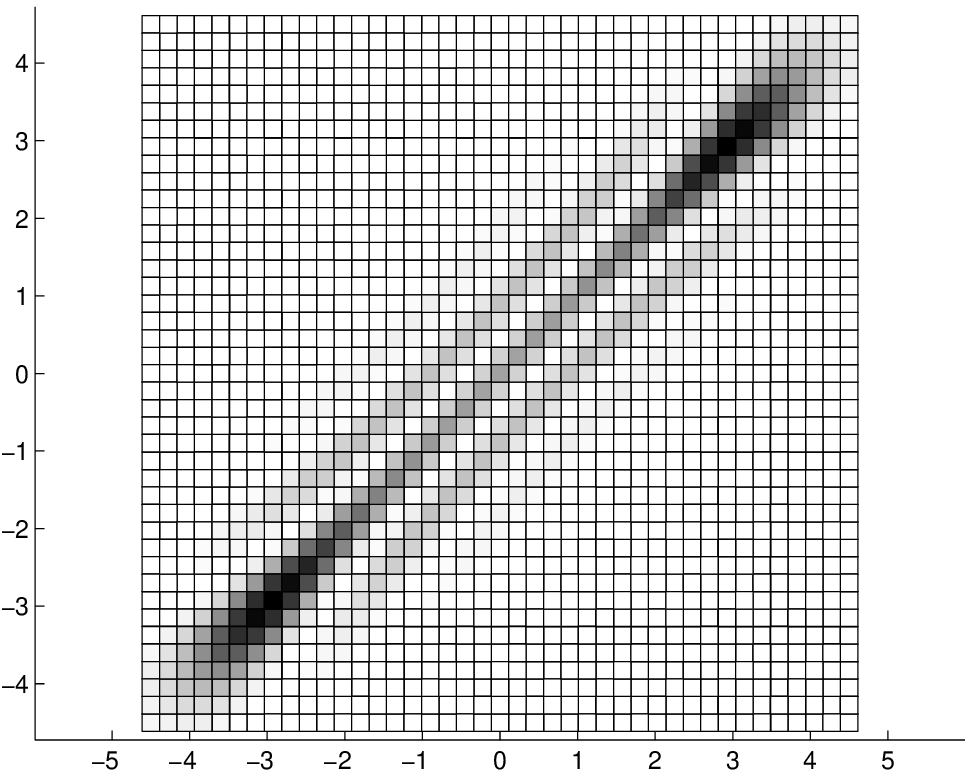}
\caption{%
Representation of the quantum prediction for the 
probability $P_{\theta,\phi}(x,y)$ 
of getting a result $x$ (horizontal axis) and $y$ (vertical axis) 
respectively upon the 
simultaneous measurements $X_{\theta}^A$ 
and $X_{\phi}^B$, where $\theta =\phi$:  
(a) $r_{0}=1.1$; (b) $r_{0}=2.5$ showing the increasing separation of peaks 
and the interference fringes characteristic of quantum superposition states. }%
\label{fig3}
\end{figure}


\begin{references}
\bibitem{16} Calculations reveal the violation of the Bell inequality to 
be sustained with only a very small ( $\Delta ^2 x \lesssim.1$) amount 
of noise (lack of resolution) 
added to the result $x$ of the quadrature phase amplitude measurement of 
$X_{\theta}^A$ (or $X_{\phi}^B$). However because of the amplification, 
prior to photodetection, due to the local oscillator $E$ this small amount of 
noise in $X_{\theta}^A$ can in principle, for sufficiently large $E$, 
translate to a large amount of noise (lack of resolution) in the detected 
photon number $s_{\theta}^A$.
\end{references}
\end{document}